# Dynamics of DNA Bubble in Viscous Medium

A. Sulaiman[1,2,4], F.P. Zen[1,4], H. Alatas[3] and L.T. Handoko[5]

[1]*Theoretical High Energy Physics and Instrumentation (THEPI), Department of Physics, Institut Teknologi Bandung, Jl. Ganesha 10, Bandung 40132, Indonesia.*
[2]*Badan Pengkajian dan Penerapan Teknologi (BPPT) ,Jl. MH. Tahmrin 8 Jakarta , Indonesia*
[3]*Theoretical Physics Division, Department of Physics, Bogor Agricultural University , Kampus Darmaga, Bogor, Indonesia*
[4]*Indonesia Center for Theoretical and Mathematical Physics (ICTMP), Jl. Ganesha 10 Bandung 40135, Indonesia.*
[5]*Group for Theoretical and Computational Physics, Research Center for Physics, Indonesian Institute of Sciences, Kompleks Puspiptek Serpong, Tangerang , Indonesia*
albertus.sulaiman@bppt.go.id

**Abstract.** The damping effect to the DNA bubble is investigated within the Peyrard-Bishop model. In the continuum limit, the dynamics of the bubble of DNA is described by the damped nonlinear Schrodinger equation and studied by means of variational method. It is shown that the propagation of solitary wave pattern is not vanishing in a non-viscous system. Inversely, the solitary wave vanishes soon as the viscous force is introduced.

## INTRODUCTION

It is well known that the transcription processes begins with separation of the double helix into a single helix called denaturation process [2]. The well known model for explaining thermal denaturation is the Peyrard-Bishop (PB) model [3]. Intensive studies on PB model have been done for example, the molecular dynamics calculation [4], including anharmonic nearest-neighbor stacking interaction [5], connection with conformation of local denaturation [9], the effects of stacking interactions [10] and the Monte-Carlo simulation [11].

The denaturation happens when the DNA is heated gradually till reaching a critical temperature. But under physiological conditions, the DNA double-helix can be spontaneously denatured locally by unzipping and fluctuating, and have a large amplitude of a localized excitations called DNA bubble [3, 4, 5, 6]. It is well known that the thermal denaturation of double strands DNA depend on the solution that surrounding the DNA molecules [7]. Effectively, the interaction of DNA with the surrounding medium coincides to viscous damping effect. It is known that, in the case of a protein-DNA system the solvating water acts as a viscous medium that makes the nucleotide oscillations to damp out [8]. Using this fact, it is necessary to take into consideration that the solving water does act as a viscous medium that could damp out DNA breathing.

The impact of viscosity was investigated by Zdrakovic et al [12, 16, 17]. The behavior of DNA dynamics in viscous solution is described by the damped nonlinear Schrodinger equation. It was shown that the observed phase diagram for the unzipping of double stranded DNA in viscous medium is much richer than the earlier suggestion theoretical work. In the present paper, we strongly rely on the impact of viscosity on the DNA dynamics especially the PB DNA model Therefore, it is necessary to improve the PB model to take into account a viscosity.

This paper discusses the viscous dissipation effect acting to DNA in the PB models. In Sec (2) formulation of the impact of a damping is derived. The continuum approximation to describe DNA breathing is described in Sec (3). Traveling wave solution based on the variational method and the discussion is given in Sec (4). The paper is ended with a summary.

## PEYRARD-BISHOP MODELWITH DAMPING EFFECT

The PB model of DNA, the motion of DNA molecules is represented by two degree of freedom $u_n$ and $v_n$ which correspond to the displacement of the base pair from their equilibrium position along the direction of the hydrogen bonds connecting the two base in pair in two different strands [3]. Making a transformation to the center of mass coordinate representing the in phase and out of phase transverse motions, $X_n = (u_n + v_n)/\sqrt{2}$ and $Y_n = (u_n - v_n)/\sqrt{2}$ respectively, the Hamiltonian of the PB model is given by [3],

$$H = \sum_n \frac{1}{2M} p_n^2 + \frac{\kappa}{2}(X_{n+1} - X_n)^2 + \frac{1}{2M} P_n^2 \\ \frac{\kappa}{2}(Y_{n+1} - Y_n)^2 + \frac{D}{2}\left(e^{-\frac{\alpha}{2} Y_n} - 1\right)^2 \quad (1)$$

where $D$ and $\alpha$ are the depth and inverse width of the potential respectively. The momentum $p_n = M dX_n/dt$, $P_n = M dY_n/dt$ and $\kappa$ is the spring constant.

As mentioned above, the studies of PB models with the viscosity was done by adding the term $-\varepsilon \gamma dY_n/dt$ in the equation of motion (EOM) [12]. In ref. [12] the nonlinear Schrodinger equation with viscous effect was solved to study the dynamics of DNA breathing. The interaction between the system with it's environment lead dissipation of energy. This means that the system is not longer conservative and reversible. The corresponding Hamiltonian formulation for dissipative system is called Caldirola-Kanai Hamiltonian in the form of a time-dependent Hamiltonian which defined as follow [13],

$$H = e^{-\gamma t} \frac{p^2}{2M} + e^{\gamma t} V(x) \quad (2)$$

where $\gamma = \eta/M$, $\eta$ is damping coefficient. The model has been used to study the quantum dissipation such as study of susceptibility for identical atoms subjected to an external force [14], coherent states for the damped harmonic oscillator [14] and dissipative tunneling of the inverted Caldirola-Kanai oscillator [15].

We propose the extension of the above approach to describe the denaturation processes in a dissipation system then the PB model with the damping effect is defined as follow,

$$H = \sum_n \frac{e^{-\gamma t}}{2M} p_n^2 + \frac{\kappa e^{\gamma t}}{2}(X_{n+1} - X_n)^2 + \frac{e^{-\gamma t}}{2M} P_n^2 \\ \frac{\kappa e^{\gamma t}}{2}(Y_{n+1} - Y_n)^2 + \frac{De^{\gamma t}}{2}\left(e^{-\frac{\alpha}{2} Y_n} - 1\right)^2 \quad (3)$$

Substituting into the Hamilton equation yields,

$$M\ddot{Y}_n + M\gamma \dot{Y}_n = \kappa(Y_{n+1} - 2Y_n + Y_{n-1}) \\ + \frac{\alpha D}{2} e^{-\frac{\alpha}{2} Y_n}\left(e^{-\frac{\alpha}{2} Y_n} - 1\right) \quad (4)$$

and

$$M\ddot{X}_n + M\gamma \dot{X}_n - \kappa(X_{n+1} - 2X_n + X_{n-1}) = 0 \quad (5)$$

The damping term $M\gamma dY_n/dt$ is similar with [12], where they add a new damping force in the EOM.

## CONTINUUM APPROXIMATION FOR STUDY DENATURATION BUBBLE

The dynamical behavior of DNA breathing can be studied by applying the continuum approximation on the equation (7). We assume that the amplitude of oscillation is small and the nucleotide oscillates around the bottom of the Morse potential but large enough due to nonlinear effect. We use the following approximation [1, 12, 16],

$$Y_n = \varepsilon \Psi_n \quad ; \quad \varepsilon \ll 1 \quad (6)$$

Substituting Eq.(6) into Eq.(3) and retaining up to the third order of Morse potential, we get

$$\ddot{\Psi}_n + \gamma \dot{\Psi}_n = \omega_0^2 (\Psi_{n-1} - \Psi_n + \Psi_{n-1}) \\ + C_m^2\left[\Psi_n + \varepsilon a_1 \Psi_n^2 + \varepsilon^2 a_2^2 \Psi_n^3\right] \quad (7)$$

where $\omega_o^2 = \kappa/M$, $C_m^2 = \alpha^2 \underline{D}/2M$, $a_1 = -3/4\alpha$, $a_2 = \alpha\sqrt{7}/24$ and $\underline{D} = 1/N \sum_n^N D_n$ is the average value of $D$. For a relatively long DNA chain, this equation can be simplified by taking full continuum limit approximation which should be valid as long as the solution under consideration changes rather slowly and smoothly along with DNA [2]. This approximation yields,

$$\frac{\partial^2 \Psi}{\partial t^2} + \gamma \frac{\partial \Psi}{\partial t} = C_0^2 \frac{\partial^2 \Psi}{\partial x^2} + C_m^2\left[\Psi + \varepsilon a_1 \Psi^2 + \varepsilon^2 a_2^2 \Psi^3\right] \quad (8)$$

where $C^2_0 = \omega^2_0 l^2$ and $l$ is a length scale. Further, we use the multiple scale expansion method, namely by expanding the associated equation into different scale and time spaces [18],

$$\frac{\partial}{\partial t} = \frac{\partial}{\partial t_0} + \varepsilon \frac{\partial}{\partial t_1} \quad ; \quad \frac{\partial}{\partial x} = \frac{\partial}{\partial x_0} + \varepsilon \frac{\partial}{\partial x_1} \quad (9)$$

By using the expansion, Eq.8 become,

$$\frac{\partial^2 \Psi}{\partial t_0^2} + 2\varepsilon \frac{\partial^2 \Psi}{\partial t_0 \partial t_1} + \varepsilon^2 \frac{\partial^2 \Psi}{\partial t_1^2} + \gamma \varepsilon \frac{\partial \Psi}{\partial t_0} + \varepsilon^2 \gamma \frac{\partial \Psi}{\partial t_1} \\ - C_0^2\left(\frac{\partial^2 \Psi}{\partial x_0^2} + 2\varepsilon \frac{\partial^2 \Psi}{\partial x_0 \partial x_1} + \varepsilon^2 \frac{\partial^2 \Psi}{\partial x_1^2}\right) \quad (10) \\ + C_m^2\left[\Psi + \varepsilon a_1 \Psi^2 + \varepsilon^2 a_2^2 \Psi^3\right] = 0$$

The zero order of the equation is an inhomogeneous wave equation and by using solution in the form $\Psi \sim \exp(i(kx_0 - \omega t_0))$ yields, $\omega^2 = C^2_m + C^2_0 k_2$. The basic behavior is the harmonic solutions, while the remaining terms lead to non-harmonic solutions, then it is reasonable to consider [3],

$$\Psi = \Psi^{(0)} + \varepsilon\Psi^{(1)} \quad (11)$$

and use the ansazt [1,3],

$$\Psi^{(0)} = \overline{\Psi}^{(1)}(x_1,t_1)e^{i\theta} + cc$$
$$\Psi^{(1)} = \overline{\Psi}^{(0)}(x_1,t_1) + \overline{\Psi}^{(2)}(x_1,t_1)e^{i2\theta} + cc \quad (12)$$

Substituting Eqs (11) and (12) into Eq (10) and the damping term can be viewed as a small perturbation due to the reason that the viscosity of water is temperature dependent, from a simple fluid mechanics argument, one can estimate the magnitude of the damping coefficient as very small at the physiological temperature [8]. The collection is not in the perturbation order but in the harmonic order i.e. $\theta$ and $2\theta$. The results are,

$$\frac{1}{2\omega}\frac{\partial^2 \overline{\Psi}^{(1)}}{\partial t_1^2} - \frac{C_0^2}{2\omega}\frac{\partial^2 \overline{\Psi}^{(1)}}{\partial x_1^2} + \frac{\gamma}{2\omega}\frac{\partial \overline{\Psi}^{(1)}}{\partial t_1} + \frac{2C_m^2}{\omega}\left|\overline{\Psi}^{(1)}\right|^2 \overline{\Psi}^{(1)}$$
$$- \frac{i}{\varepsilon}\left(\frac{\partial \overline{\Psi}^{(1)}}{\partial t_1} + V_g \frac{\partial \overline{\Psi}^{(1)}}{\partial x_1}\right) = 0 \quad (13)$$

where $V_g = C_0^2 k/\omega$. It's convenience to write the equation in traveling wave coordinate as,

$$\xi = x_1 - V_g t_1 \quad ; \quad \tau = \varepsilon t_1 \quad (14)$$

And by assuming the damping factor only depend on time coordinate then the equation become,

$$i\frac{\partial \overline{\Psi}^{(1)}}{\partial \tau} + \Lambda_1 \frac{\partial^2 \overline{\Psi}^{(1)}}{\partial \xi^2} + \Lambda_2 \frac{\partial \overline{\Psi}^{(1)}}{\partial \tau} + \Lambda_3 \left|\overline{\Psi}^{(1)}\right|^2 \overline{\Psi}^{(1)} = 0$$
.................(15)

where $\varepsilon$ where we have ignored the second order, $\Lambda_1 = C_0^2 C_m^2/(2\omega^3)$, $\Lambda_2 = \gamma\varepsilon/(2\omega)$ and $\Lambda_3 = 2C_m^2/\omega$. The equation is called the Damped Nonlinear Schrodinger Equation (DNLS) that also obtained in Zdrakovic et al 2005 [12].

## TRAVELING WAVE SOLUTION

Let us solve Eq (15) in term traveling wave by variational methods based on the Lagrangian formulation. It is well know that for the case with $\gamma = 0$, the Nonlinear Schrodinger equation admits the following traveling wave solution [2, 12, 16],

$$\overline{\Psi}^{(1)} = A_0 \sec h\left[\tfrac{1}{L}(\xi - u_e\tau)\right]e^{-i(\tilde{k}\xi - \tilde{\omega}\tau)} \quad (16)$$
where $A_0 = \sqrt{(u_e^2 - 2u_e u_c)/(2\Lambda_1 \Lambda_3)}$, $L = \sqrt{2} \Lambda_1 /\sqrt{(u2e}$ $-2u_e u^c)$, $\bar{k} = u_e^2 \Lambda_1$, $\tilde{\omega} = u_e u^c/2 \Lambda_1$ and . Here, $u_e$ is the envelope wave velocity and $u_c$ is the carrier wave velocity, satisfying $u_e^2 - 2u_e u^c > 0$. Based on the corresponding variational methods to solve the damped-forced nonlinear Schroedinger equation, one can use the solution (16) as the related basic form and considering its amplitude, width, phase velocity and the position of the soliton to be time dependent [18, 19, 20]. Let us write the 1-soliton in the following form,

$$\overline{\Psi}^{(1)} = \eta(\tau)\sec h[\eta(\tau)(\xi - \zeta(\tau))]e^{-i(\theta(\tau)\xi - \phi(\tau))} \quad (17)$$

The dynamics of $\eta$, $\theta$, $\zeta$ and $\phi$ function can be obtained by using the variational methods. The Lagrangian for the DNLS is given by,

$$L = \int \tfrac{i}{2}\left(\overline{\Psi}^{(1)}_\tau \overline{\Psi}^{(1)*} - \overline{\Psi}^{(1)*}_\tau \overline{\Psi}^{(1)}\right) - \Lambda_1\left|\overline{\Psi}^{(1)}_\xi\right|^2$$
$$+ \Lambda_3\left|\overline{\Psi}^{(1)}\right|^4 + \tfrac{\Lambda_2}{2}\left(\overline{\Psi}^{(1)}_\tau \overline{\Psi}^{(1)*} - \overline{\Psi}^{(1)*}_\tau \overline{\Psi}^{(1)}\right)d\xi \quad (18)$$

Substituting Eq.(17) into Eq. (18) and using $\int \text{sech}(a\xi)d\xi = \pi/a$ and $\int \text{sech}2(a\xi) \tanh(a\xi)d\xi = 0$ yield,

$$L = 2\eta\zeta\dot\theta + \eta\dot\phi + i2\Lambda_2\eta\zeta\dot\theta$$
$$+ i\Lambda_2\eta\dot\phi + 4\Lambda_1\eta\theta^2 + \tfrac{4}{3}\Lambda_3\eta^3 \quad (19)$$

The EOM can be easily obtained by solving the Euler-Lagrange equation. This yields,

$$2(1+i\Lambda_2)\eta\dot\zeta + 2(1+i\Lambda_2)\zeta\dot\eta = 8\Lambda_1\eta\theta$$
$$2(1+i\Lambda_2)\zeta\dot\theta + (1+i\Lambda_2)\dot\phi = 4(\Lambda_1\theta^2 - \Lambda_3\eta^2) \quad (20)$$
$$(1+i\Lambda_2)\dot\eta = 0 \quad ; \quad (1+i\Lambda_2)\dot\theta = 0$$

The solution of the last two terms in Eq (20) are $\eta = \eta_0$ and $\theta = \theta_0$ respectively. It's not difficult to show that the solution of Eq.20 is given by,

$$\zeta(\tau) = \Delta_\xi(1 - i\Lambda_2)\tau + \zeta_0$$
$$\phi(\tau) = \Delta_\phi(1 - i\Lambda_2)\tau + \phi_0 \quad (21)$$

where $\Delta\zeta = (4\Lambda_1\eta_0\theta_0)/(1 - \Lambda_2^2)$ and $\Delta\phi = 4(\Lambda_1\theta_0^2 - \Lambda_3\eta_0^2)/(1 - \Lambda_2^2)$. Then the explicit solution is given by,

$$\overline{\Psi}^{(1)} = \eta_0 \sec h[\eta_0(\xi - \Delta_\xi(1 - i\Lambda_2)\tau)]$$
$$\times \exp[-i(\theta_0\xi - \Delta_\phi(1 - i\Lambda_2)\tau + \phi_0)] \quad (22)$$

The profile of denaturation bubble is obtained by using Eqs (11) and (12). Then we obtain the soliton solution as follow,

$$\Psi(x,t) = 2\Psi^{(1)}\cos(kx - \omega_0 t) +$$
$$\varepsilon\left|\Psi^{(1)}\right|^2[3 + \cos(2(kx - \omega_0 t))] \quad (23)$$

We simulate the solution for the model parameter given by κ = 8*Nm*, *M* = 5.1× 10$^{-25}$*kg*, α = 2 ×10$^{10}$*m*$^{-1}$, *D* = 0.1*eV* and *l* = 3.4×10$^{-10}$*m* is length scale [12]. The system of unit (*A°*,*eV*) defines a time unit (*t.u.*) equal to 1.021×10$^{-14}$*s* [17]. The simulation result is depicted in Fig.1. The homogeneous solution (original PB model) demonstrates a sort of a modulated solitonic wave where the hyperbolic and cosine terms correspond to the wave number of the envelope and the carrier wave respectively. The DNA breathing with damping factor propagates faster than original one, but disperse and finally die out. From the figure we observe that the amplitude of the soliton structures decreases as time progresses, because of the damping due to viscosity of the surrounding medium. Therefore, when the damping is high, the wave patterns are expected to travel faster only for a short time and will vanish. These results are in coincidence with previous work [12]. It is interested when external force applied to the system. The work is still in progress.

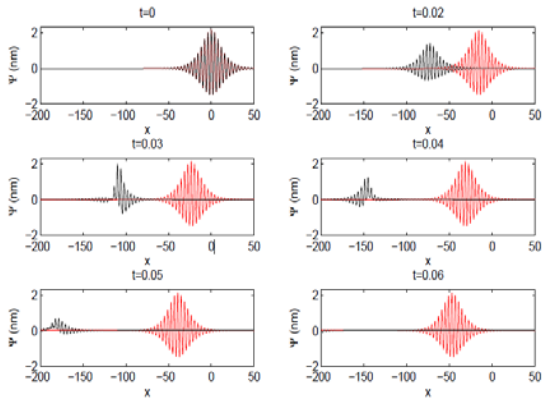

**FIGURE 1.** The DNA breathing in the second case where *x* denotes the continuum base pair in the present model (*black line*) and the original PB model (*red line*) with γ = 0.05 .

## SUMMARY

The impact of viscous fluid to the PB DNA breathing is investigated. We have proposed a PB model with the damping effect which is described by an extended time-dependent Caldirola-Kanai Hamiltonian. Taking full continuum approximation and using the multiple scale expansion method, the EOM is nothing else than the DNLS equation. We have shown that when the viscous forces are not taken into account, the system reduces to the ordinary NLS equation. When the viscosity, due to the medium which surrounds the molecule, is taken into account, the amplitude of base pair oscillations is described by the DNLS equation. It is shown that the propagation of solitary wave pattern is not vanishing in a non-viscous system. Inversely, the solitary wave vanishes soon as the viscous force is introduced.

## ACKNOWLEDGMENTS
AS thank Program Insentif GOSAT PTISDA-BPPT for financial funding. This work also is funded by the Indonesia Ministry of Research and Technology and the Riset Kompetitif LIPI in fiscal year 2011 under Contract no. 11.04/SK/KPPI/II/2011. FPZ thanks Research KK ITB 2011.